\documentclass[conference]{IEEEtran}
\IEEEoverridecommandlockouts
\usepackage{cite}
\usepackage{amsmath,amssymb,amsfonts}
\usepackage{algorithmic}
\usepackage{graphicx}
\usepackage{textcomp}
\usepackage{xcolor}
\usepackage{multirow}

\def\BibTeX{{\rm B\kern-.05em{\sc i\kern-.025em b}\kern-.08em
T\kern-.1667em\lower.7ex\hbox{E}\kern-.125emX}}

\begin{document}

\def\x{{\mathbf x}}
\def\L{{\cal L}}

\makeatletter
    \newcommand{\linebreakand}{%
      \end{@IEEEauthorhalign}
      \hfill\mbox{}\par
      \mbox{}\hfill\begin{@IEEEauthorhalign}
    }
    \makeatother
\title{Matrix Syncer - A Multi-chain Data Aggregator For Supporting Blockchain-based Metaverses}

\author{\IEEEauthorblockN{1\textsuperscript{st} Xinyao Sun}
\IEEEauthorblockA{\textit{Matrix Labs Inc.} \\
asun@matrixlabs.org}
\and
\IEEEauthorblockN{2\textsuperscript{nd} Yi Lu}
\IEEEauthorblockA{\textit{Dapper Labs} \\
amberluyi@dapperlabs.team}
\and
\IEEEauthorblockN{3\textsuperscript{rd} Jinghan Sun}
\IEEEauthorblockA{\textit{Chinese University of HongKong, Shenzhen} \\
jinghansun@link.cuhk.edu.cn}
\and
\linebreakand 
\IEEEauthorblockN{4\textsuperscript{th} Bohao Tang}
\IEEEauthorblockA{\textit{Dapper Labs} \\
bohao.tang@dapperlabs.team}
\and
\IEEEauthorblockN{5\textsuperscript{th} Kyle D. Rehak}
\IEEEauthorblockA{\textit{Matrix Labs Inc.} \\
kyle@matrixlabs.org}
\and
\IEEEauthorblockN{6\textsuperscript{th} Shuyi Zhang}
\IEEEauthorblockA{\textit{White Matrix Tech Ltd.} \\
tim@whitematrix.io}
}

\maketitle

\begin{abstract}
Due to the rising complexity of the metaverse's business logic and the low-latency nature of the metaverse, developers typically encounter the challenge of effectively reading, writing, and retrieving historical on-chain data in order to facilitate their functional implementations at scale. While it is true that accessing blockchain states is simple, more advanced real-world operations such as search, aggregation, and conditional filtering are not available when interacting directly with blockchain networks, particularly when dealing with requirements for on-chain event reflection. We offer Matrix Syncer, the ultimate middleware that bridges the data access gap between blockchains and end-user applications. Matrix Syncer is designed to facilitate the consolidation of on-chain information into a distributed data warehouse while also enabling customized on-chain state transformation for a scalable storage, access, and retrieval. It offers a unified layer for both on- and off-chain state, as well as a fast and flexible atomic query. Matrix Syncer is easily incorporated into any infrastructure to aggregate data from various blockchains concurrently, such as Ethereum and Flow. The system has been deployed to support several metaverse projects with a total value of more than \$15 million USD.
\end{abstract}
\section{Introduction}
\label{sec:intro}
Metaverse is a portmanteau of the prefix "meta" (which means "beyond") and the suffix "verse" (shorthand for "universe"). It is derived from Neal Stephenson's science fiction novel Snow Crash \cite{hawking1988} and literally refers to a cosmos beyond our physical world. As stated in \cite{dionisio20133d},  the metaverse 
is predicated on advancements in four key areas: immersive realism, ubiquitous access and identification, interoperability, and scalability. To enable the construction of a completely functional metaverse, each of those areas must be adequately developed. The development can be grouped into several sub-domains, including numerous multimedia technologies -  network transmission and prototyping, computer graphics, image processing, virtual reality, and augmented reality \cite{dionisio20133d}.


Globalization has increased the volume of international communication and cooperation on a global scale, however geographic distance is an objective hindrance that increases costs. Additionally, as a result of the COVID-19 pandemic, many events have been suspended to comply with pandemic preventive standards \cite{han2021investigation}. 
These stringent requirements have created a major opportunity for initiatives including teleconferences and virtual gatherings, in which the metaverse could provide significant accessibility to meet those social requirements \cite{han2021investigation}. Moreover, decentralization has been described as a critical component of initiating the fifth phase of metaverse development \cite{dionisio20133d}. Decentralized development has resulted in the decoupling of the client and server sides of a virtual world system, which the blockchain protocol has facilitated in recent years. The metaverse is expected to connect everyone on the entire globe.
Study \cite{duan2021metaverse} asserted that blockchain technology is critical for ensuring the sustainability of metaverse ecosystems by ensuring decentralization and fairness. 

From a macro perspective, a three-layer metaverse architecture is outlined in \cite{han2021investigation} as 1) ecosystem, 2) interaction, and 3) infrastructure. Each layer is composed of distinct modules that work together to make a comprehensive, interactive, and functional metaverse. As a result, having an interoperable, robust, and low-latency middleware between each module becomes critical for ensuring that diverse virtual worlds are able to seamlessly connect and overlap. 
As we progress toward a more decentralized metaverse, we must examine the solutions we wish to emphasize in the metaverse and other advancements. 


\section{Related Works}
\label{sec:related_works}
\subsection{Traditional Web 2.0 Metaverse}
Numerous companies, game creators (\textit{Roblox}\footnote{https://www.roblox.com/} and \textit{Epic Games}\footnote{https://www.epicgames.com/}), software giants (Microsoft, Amazon), social media conglomerates (Facebook - now Meta, Twitter), and graphics processor manufacturers (Nvidia) are involved in the metaverse, beginning with online events. Extensive research and development has been conducted to optimize virtual gathering in the metaverse; NVIDIA's Omnivers\footnote{https://developer.nvidia.com/nvidia-omniverse-platform} is a scalable, multi-GPU real-time reference development platform for 3D simulation and design collaboration; and Alibaba's \textit{Cloud Metaverse}\footnote{https://www.alibabacloud.com/solutions/metaverse} has been released for the purpose of utilizing cloud computing to construct the entire virtual world as a service. Meta recently launched  \textit{Horizon Worlds}\footnote{https://www.oculus.com/facebook-horizon/}, a virtual-reality social networking platform that allows up to 20 avatars to explore, hang out, and build in the virtual realm, as well as a number of revolutionary gadgets, controllers, and supporting hardware, including VR gloves with haptic feedback. However, all of these advancements in multimedia technology are web 2.0 based. The service providers are centralized identity providers, and users' digital identities are produced and held centrally. These advancements are not sufficient to create a transparent, stable, and sustainable digital economy, where digital properties belong to the users, not the operators \cite{ryskeldiev2018distributed}.

\subsection{Blockchain-based Web 3.0 Metaverse}
The success of Bitcoin \cite{nakamoto2008bitcoin}, 
as a decentralized transaction system has garnered significant attention. Later, in 2013, Vitalik Buterin proposed Ethereum \cite{buterin2014next}, a decentralized computation platform that introduced the use of smart contracts to execute programs autonomously and transparently on the blockchain. Since then, a variety of different public blockchain networks have been established, including \textit{Flow}\footnote{https://www.onflow.org/}, \textit{EOS}\footnote{https://eos.io/}, etc. each of which supports the development of decentralized applications (DApps) and has its own design philosophy aimed at improving the user experience and system performance. Numerous DApp-based metaverses, such as \textit{Sandbox}\footnote{https://www.sandbox.game} and  \textit{Decentraland}\footnote{https://decentraland.org/}, have attracted increasing attention and consumers in recent years, resulting in significant revenues \cite{han2021investigation}. This demonstrates how decentralization's power could ensure that digital properties are unique, permanent, and transferable, which %
benefits the metaverse's development and enables the construction of a fair, free, and sustainable society \cite{harwick2020s}. Web 2.0's inadequacies, along with the existence of public blockchain technology, have gradually increased public awareness of privacy, data rights, censorship, and identity difficulties. These factors have facilitated the transition of users to a more decentralized Web 3.0 metaverse.

\subsection{Indexing Blockchain Data}
We are still in the early stages of the development of decentralized technologies. When developing a decentralized application for the metaverse, the developer frequently encounters constraints on on-chain compute power and storage, along with a wide variety of public blockchain interfaces \cite{liu2021make}. 
Due to the increasing complexity of the metaverse's business logic and the requirements of low-latency user experiences, properly reading, writing, and retrieving historical on-chain data in order to support their functional implementation at scale has always been an inevitable challenge \cite{farmer2021decentralized}. The \textit{Graph}\footnote{https://thegraph.com/} is a popular decentralized protocol for indexing and querying data on Ethereum-based blockchains. It enables the querying of data that was inaccessible directly before. However, its support provides only a restricted set of interfaces for transforming data from an on-chain structure to a GraphQL-compatible schema. It offers fast and efficient querying of historical blockchain data, but does not address reflective requirements such as making an external service request or initiating transactions in response to on-chain states and received events.

\begin{figure*}[ht!]
    \centering
    \includegraphics[width=0.8\textwidth]{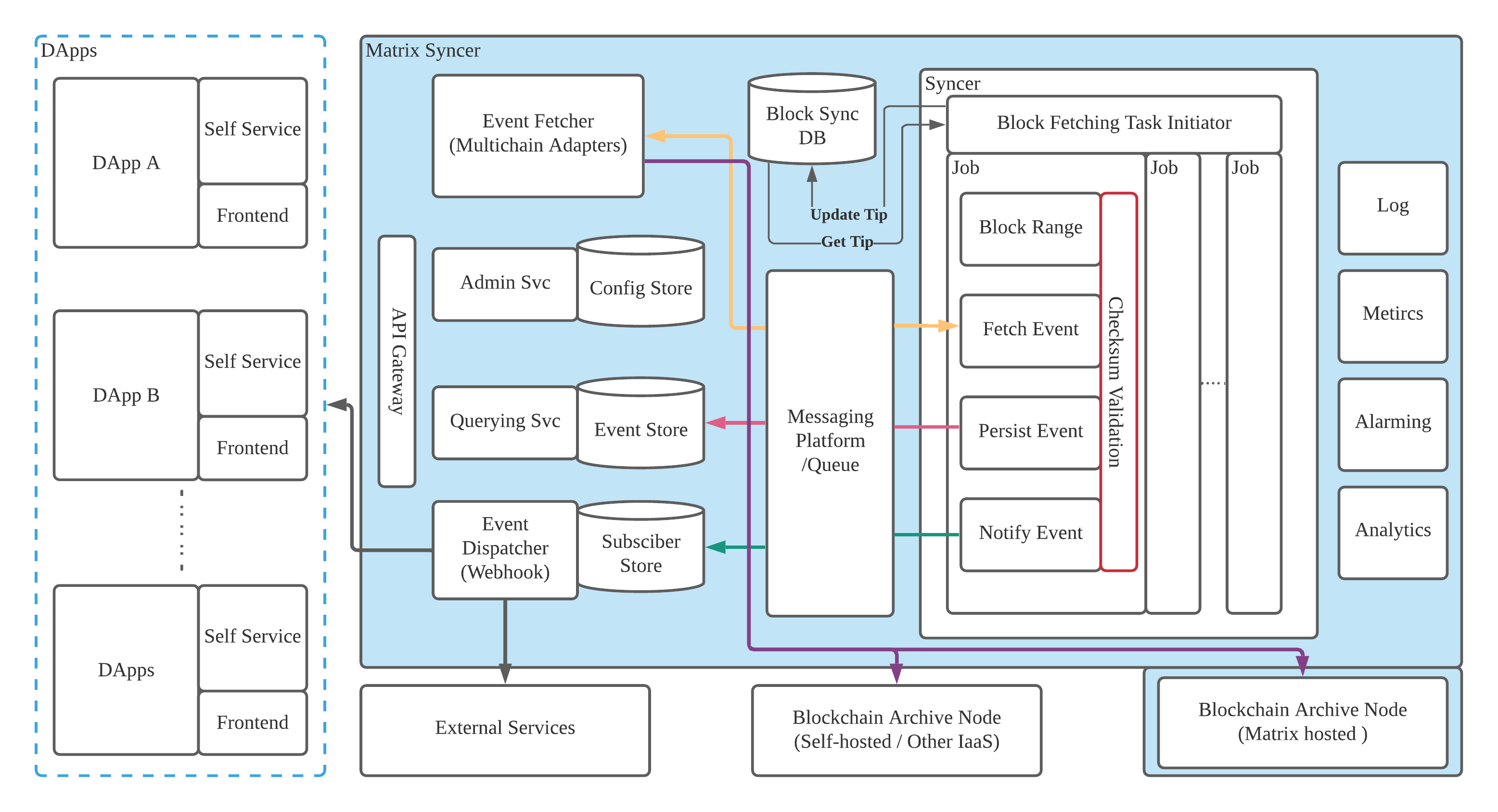}
    \caption{An overview of the Matrix Syncer architecture}
    \label{fig:architecture}
\end{figure*}

\subsection{Web3 Infrastructure as a Service}
Typically, developers can establish their own web 3.0 infrastructure and communicate directly with a self-hosted blockchain node in order to fetch the on-chain state change (event) and perform reflective processing on their own services. Unfortunately, it is widely known that setting up a self-hosted node is costly and time-consuming \cite{singh2018blockchain}.
There is a demand for utilities that decrease the entrance barrier and make blockchain data more accessible. Infrastructure-as-a-Service (IaaS) offerings are among the most critical. \textit{Infura}\footnote{https://infura.io/} is leading the charge, providing developers, decentralized application teams, and corporations from a variety of industries with a suite of tools for connecting their apps to blockchain networks. 
\textit{Alchemy}\footnote{https://www.alchemy.com/} further extends Infura by adding support for Ethereum layer-2 and other public blockchains (e.g., Flow). The most significant ultility they have introduced is the Alchemy Notify API, which uses webhooks to trigger external actions, allowing for on-chain reflective implementations. However, neither of these projects offer a proper way to retrieve historical on-chain events through traditional query requests.

\subsection{Gap of Research and Development }
Top traditional IT businesses are attempting to improve the virtual social gathering user experience in order to make it a more effective and secure new way of living. 
To meet the demands of a decentralized metaverse, developers will need to overcome some key issues caused by current web 3.0 limitations. Numerous products and services have been developed to facilitate the development and deployment of smart contracts in the modern era, such as ChainIDE \cite{wu2020chainide}. One of the key constraints on the scalability and accessibility of the majority of decentralized projects is the inability to index and react effectively to on-chain events that trigger self-business logic after the contracts have been deployed. Furthermore, blockchain is currently a niche industry, with one of the key causes being that blockchains lack interoperable infrastructure, which prevents applications from being deployed on a large scale \cite{liu2021make}. Thus, a solid infrastructure that connects web 2.0 and web 3.0 development stacks and supports multiple blockchains is essential to accelerate the decentralized metaverse revolution.

In this work, we present Matrix Syncer, the ultimate middleware that bridges the data access and event trigger gaps between blockchains and end-user services and applications. Matrix Syncer is intended to make it easier to consolidate on-chain data into a distributed data warehouse while also enabling customizable on-chain state transformation for scalable storage, access, and retrieval. It provides a uniform layer for both on-chain and off-chain states, as well as quick and flexible atomic querying and event triggering. Furthermore, the Matrix Syncer can easily be  integrated into any infrastructure to aggregate data from multiple blockchains simultaneously.

\section{Proposed Framework}
\label{sec:proposed_frameworks}
\subsection{System Design}
The metaverse, we believe, is an event-driven environment in which each user interaction should trigger a chain of reflections altering the state of related properties. Such events are typically required by an application in the metaverse to execute reactive activities 
in response to the current world state and system logic. Matrix Syncer is a cloud-based IaaS platform that allows developers to effortlessly develop the aforementioned workflows without having to worry about blockchain-related setups or maintenance. The overview of the system architecture is shown in Fig. \ref{fig:architecture}. Matrix Syncer can help developers to create web 3.0 compatible applications while maintaining their familiar web 2.0 development style. 

\subsection{Event Registration}
As seen in the figure, developers can configure their EOIs by registering them in the Block Sync DB, which registers the EOI's global unique identity derived from the chain type, contract address, and event signature. Each event has its own set of variables that control how the block syncer distributes jobs; the three most critical ones are \textit{initBlockHeight}, \textit{syncedStartBlockHeight}, and \textit{syncedLatestBlockHeight}. Developers define the \textit{initBlockHeight} when registering a new event. It serves as the starting point for synchronizing a specific event, which is typically the block height of the contract deployment. This is necessary because the contract may have previously been deployed and performed transactions before enrolling on our platform.  Here, \textit{syncedStartBlockHeight} refers to the block height at the time this event was first synced, and \textit{syncedLatestBlockHeight} refers to the block height that the block syncer scanned for this event the most recently.

\subsection{Block Synchronization}
The block syncer distributes \textbf{regular} synchronization jobs in parallel for each event with the batch size $K = min(mB, cL - \gamma - syncedLatestBlockHeight)$, where $mB$ denotes the maximum batch size for a given chain and $cL$ indicates the most recently minted block height from connected archive node. $mB$ is an adaptive parameter that can be adjusted based on the transaction per section (TPS) rate as well as the rate of finalized blocks in order to ensure that the processing capacity of the entire system is sufficient for each chain's throughput.
$\gamma$ is a chain-specific parameter that prevents the most recent confirmed blocks from being reverted, as is the case with Ethereum when an archive node receives an uncle block and later reverts it to adhere to the longest chain protocol \cite{ritz2018impact}. As a result, $\gamma$ can be set to a number (e.g., 5 for Ethereum, it is the default number of blocks for confirmation of freshly minted blocks defined by the Go Ethereum client) to ensure that the synced block information is already persistent in the network. When \textit{initBlockHeight} is less than \textit{syncedStartBlockHeight}, Block Syncer will distribute a dedicated group of \textbf{backfilling} jobs to rapidly scan for the target event between [\textit{initBlockHeight}, \textit{syncedStartBlockHeight}] in parallel. Once the \textbf{backfilling} task has been completed, the \textit{syncedStartBlockHeight} of this event will be set to equal to the \textit{initBlockHeight}.

\subsection{Event Fetching}
The fetcher is implemented using corresponding interfaces according to different blockchains. It reads all EOIs for a certain block range and converts them to a meaningful data structure that can be stored in a database. The majority of public blockchains provide an interface or SDK for interacting with their nodes in order to perform simple chain state or event queries. For example, on Ethereum, we can use \textit{ether.js}\footnote{https://docs.ethers.io/}to fetch events by block range from an archive node. It's a little more complicated when dealing with Flow since Flow's blockchain data is segmented via Sporks. Each spork stores data for a specific block range. When we use the official Flow endpoint, we can only request blocks or events from the current Spork. To address these issues, we designed the ultimate event fetcher as a middleware module that automatically subdivides large block ranges into smaller sections with associated Spork endpoints and then merges all fetched events together to provide a level of usability similar to Ethereum's. The source code is available at [GitHub]. (reveal later to respect to the double-blind rule). In our instance, we just need to construct an adapter for each chain using a suitable library and connect to an archive node capable of retrieving historical data. The user can specify the type of nodes they wish to employ. They can either create their own self-hosted node or provide the endpoint for third-party IaaS platforms such as Infura or Alchemy. Additionally, Matrix Syncer offers its own archive node for multiple chains, providing developers more freedom for cost-effectiveness optimization.

\subsection{Event Persistence and Indexing}
When registering an event, developers have the option of specifying a database scheme for converting on-chain events into a database (event store). All event fetchers will feed EOIs into a persistent module as decoded data. The persistent module converts feed-in events into various data structures defined by the developers and saves them in the event store. All subsequent queries will be performed at the data model specified in the schema. We adopted \textit{DynamoDB}\footnote{https://aws.amazon.com/dynamodb/} as a cloud database solution to achieve 1) horizontal scaling via managed partitioning and sharding, 2) high availability with assured SLAs, and 3) high-level consistency. After the backfilling process is completed, developers will be able to do advanced queries, such as filtering, pagination, sorting, grouping, and joining result sets. Matrix Syncer is designed to accommodate multiple chains. Hence, it brings the important advantage of allowing a unified data structure for DApps, which support similar business logic on multiple chains. Users can define the same schema for different chains in order to convert heterogeneous raw events into a standard format and facilitate interoperability in the metaverse.

\subsection{Event Data Integrity Assurance}
We assure end-to-end data completeness for each block fetching task by performing a checksum verification at the end of each step. During the event fetching and persistence step, we calculate the number of persisted events grouped by registered event type and the number of non-persisted events. The checksum equation is:
\begin{multline}
 count(allEvents) \iff count(nonPersistedEvents) \\ \sum_{type}count(persistedEventsPerType)   
\end{multline}
If this equation holds, we persisted all registered events, and non-registered events are correctly skipped. In the Nofity Event phase, we will check if the $count(notificationSent)$ equals to $\sum_{type}  count(persistedEventsPerType)$, which is the checksum we calculated in the previous step to make sure no registered events are missing during the notification step. These checksums are persisted in Block Sync DB as structured data and can be used for analytic and monitoring purposes. For instance, any checksum verification failure will result in an instant alarm. Developers can then use the persisted checksums of each step to expedite debugging and root cause analysis.

\subsection{Reflective Hook and Event Queue}
To build a fault-tolerant robust system, we employ \textit{Apache Kafka}\footnote{https://kafka.apache.org/} on the side to manage inter-module communication and jobs. It permits asynchronous communication, which optimizes the data flow throughout the system. Queues make our intermediate event data persistent, which improves reliability and reduces errors when different pieces of our system are unavailable. Additionally, it can be quickly scaled to distribute workload among a fleet of users during peak periods. 
Once the event has been passed from the fetcher to the event store, if there are registered subscribers for a synchronized EOI, the event dispatcher will notify them of the structured event information via a webhook. Webhooks enable users to be notified when an EOI occurs on the blockchain. Rather than querying the server continuously to see whether the state has changed, webhooks deliver information as it becomes available, which is far more efficient and advantageous for developers. Webhooks operate by registering a URL endpoint to which notifications should be sent when specified ROI occurs. The developer maintains complete control over the endpoint, which could be a third-party service or one of their own. 

\begin{table}[]
\label{tb:snap}
\caption{Market statistics of supported metaverse projects}
\resizebox{\columnwidth}{!}{%
\begin{tabular}{|c|c|c|c|l|c|}
\hline
\textbf{Project}                               & \textbf{Blockchain} & \textbf{\# Tokens} & \textbf{\# Trans} & \textbf{\begin{tabular}[c]{@{}l@{}}\$ Init\\ Sales\end{tabular}} & \textbf{\begin{tabular}[c]{@{}c@{}}\$ Volumn\\ Traded\end{tabular}} \\ \hline
\textit{\textbf{RiverMen}}                     & Ethereum            & 10.0K              & 3327              & 1.6M                                                             & 4.5M                                                                \\ \hline
\multirow{2}{*}{\textit{\textbf{MatrixWorld}}} & Ethereum            & 1.8K               & 1416              & 5M                                                               & 4M                                                                  \\ \cline{2-6} 
                                               & Flow                & 2.0K               & 1518              & 5M                                                               & 1.5M                                                                \\ \hline
\end{tabular}%
}

\end{table}
\section{Operation Cases and Statistics}
\label{sec:evaluation}
Matrix Syncer has been deployed in the industry to serve a variety of decentralized metaverse projects, we'll look at two of them here: \textit{RiverMen}\footnote{https://www.rivermen.io/} and  \textit{MatrixWorld}\footnote{https://matrixworld.org/}, which are both representative projects with good operation cases. Table \ref{tb:snap} provides an overview of their market statistics. Rivermen is known as the world's first metaverse project dedicated to exporting traditional Chinese culture and first released in August. MatrixWorld is our first-party project, which was launched in October and is the first multi-chain support metaverse project. It currently supports the Ethereum and Flow networks and is among the top three projects by transaction volume (unofficial) on the Flow network.

\begin{figure}[ht]
    \centering
    \includegraphics[width=\columnwidth]{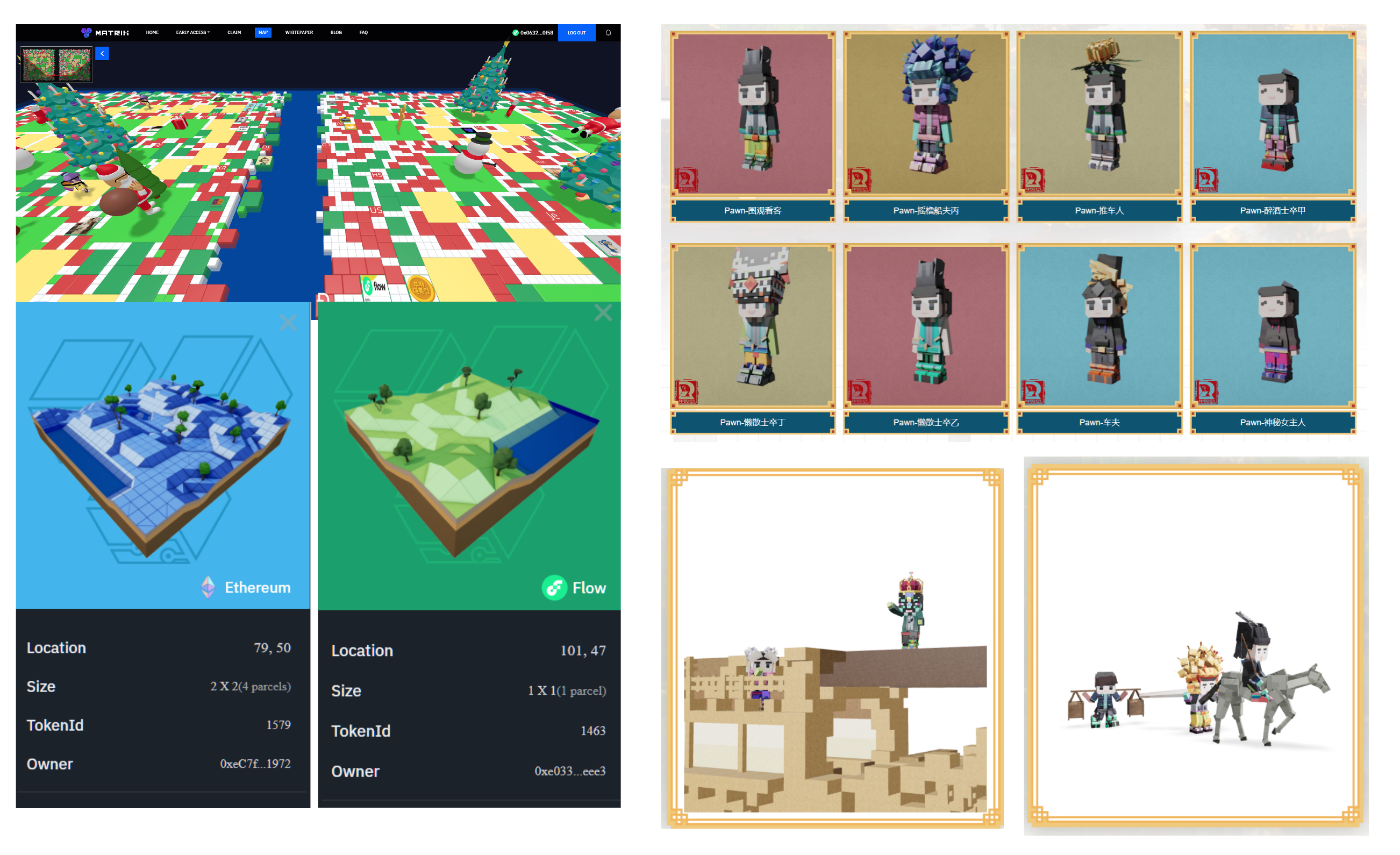}
    \vspace{-0.6cm}
    \caption{Left: MatrixWorld project, where users can interact their lands which are minted on Ethereum and Flow blockchains, Right: RiverMen pawn tokens (above) can be used to mint new RiverSpace Token (below).}
    \label{fig:projects}
\end{figure}

\subsection{Event Data Indexing}
It is commonly established that dApps must provide a front-end user interface as shown in Figure \ref{fig:projects} in order to provide an interactive user experience. RiverMen's website has a feature to display users' own tokens and perform advanced queries among their tokens as well as the entire collection. The typical smart contract's interface is insufficient for those scenarios. However, by defining a mapping scheme in Matrix Syncer, all events emitted by minting and transferring are well tracked in the event store with a timestamp. To meet the aforementioned needs, RiverMen's front-end can easily call query APIs to our event query service endpoint. Additionally, because all events are timestamped when they are saved to the store, developers can perform advanced queries, such as retrieving the entire transaction history of a particular token or obtaining stats such as the number of transactions of a given token within a specified time window. Similarly, in MatrixWorld, users can locate and investigate all land information directly from the map interface \ref{fig:projects} (left). We defined minting and transfer mapping schemas for the Ethereum and Flow blockchains and unified their data structures in the event store. Later on, our front-end can simply query the needed information without having to write two data parsers for the distinct chains. We demonstrated that Matrix Syncer platform can significantly reduce the effort required to onboard a new project with historical on-chain data querying requirements, regardless of the blockchain it is built on or the number of chains it supports concurrently.

\subsection{Event Reflection}
Apart from advanced on-chain state queries, event reflection processing is critical for many products with more complicated business logic. RiverMen introduced the \textit{RiverSpace}\footnote{https://opensea.io/collection/river-space} token, which is capable of minting new tokens through the fusion of a set of RiverMen tokens as shown in Figure \ref{fig:projects} (right). The 3D model of the newly created RiverSpace token is built and rendered dynamically based on the metadata of the RiverMen tokens. However, due to the restricted compute capacity on-chain, these procedures must be performed off-chain. Here, the RiverMen team registered an event triggered by the RiverSpace contract that contained information about the token IDs of each RiverMen token used to fuse the newly minted RiverSpace token. By registering the target event and the webhook on Matrix Syncer, they only need to develop rendering and metadata services to handle off-chain processing. Once their services receive the event, they will begin rendering the new 3D model based on the metadata of the component tokens, which can be fetched through their token IDs. Once the rendering process is complete, they can update their own metadata services to enable the front end to display RiverSpace 3D mode. Here, developers are able to concentrate entirely on their own business logic without having to implement or maintain any of the servers or infrastructure required to interact with blockchains.

\subsection{Monitoring and Analytics}

Following the best practices of web services development, Matrix Syncer uses \textit{GPL}\footnote{https://www.opencue.io/docs/other-guides/monitoring-with-prometheus-loki-and-grafana/} (Grafana/Prometheus/Loki) stack for metrics, alarms, and log viewer for regular DevOps purposes. Aside from that, we also utilize our persisted checksums for real-time alarming and analytics. We set up instant alarms during the Block Fetching Job execution and stream the checksum data to Grafana dashboard to have double insurance. In MatrixWorld, the checksum failure alarm helped us discover that the data sync error was due to an unstable data node provider. In RiverMen, the checksum stats mismatch helped us find a bug in the event processing chain. Furthermore, we ran an analytics query on our persisted checksum stats to help our customers and our team understand each typed event's processing rate and distribution over time. 

\begin{figure}[ht]
    \centering
    \includegraphics[width=\columnwidth]{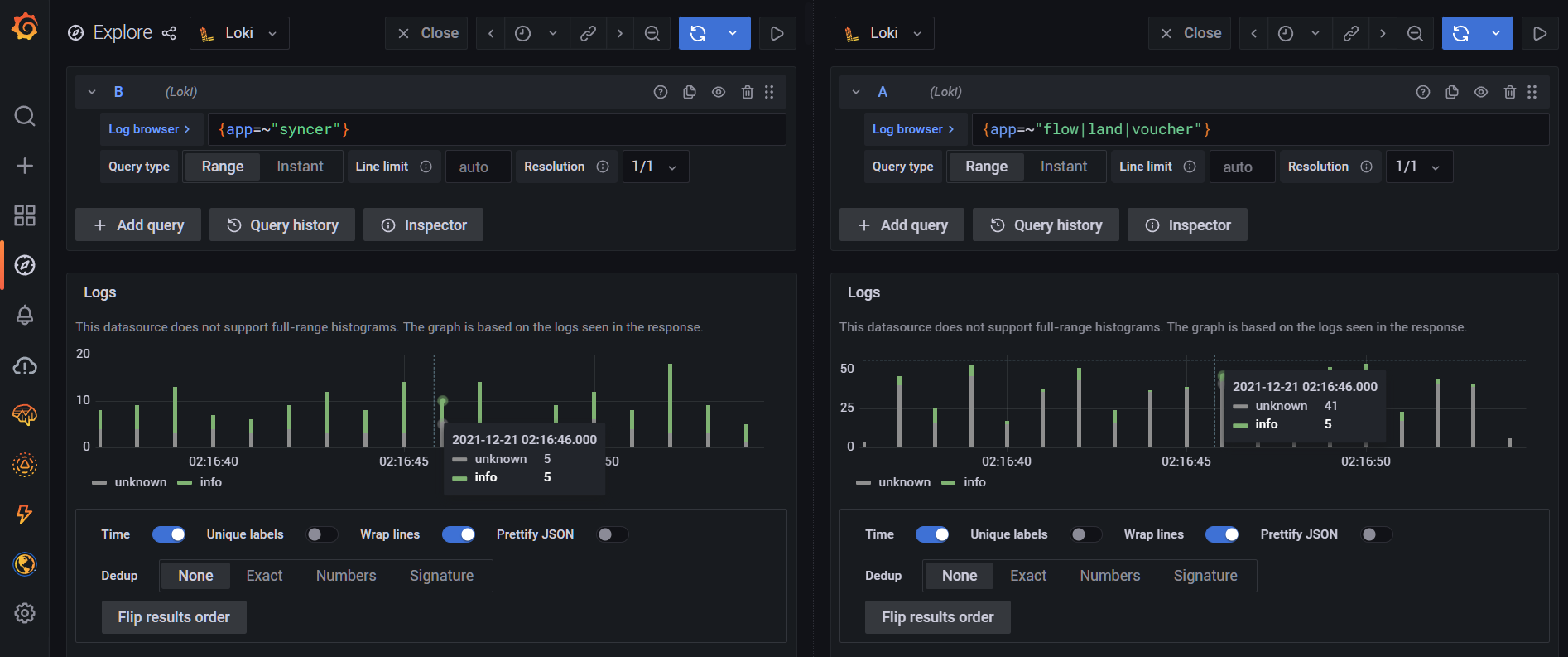}
    \caption{Monitoring dashboard and data analytics with Grafana}
    \label{fig:dashboard}
\end{figure}

\section{Conclusions}
\label{sec:conclusions}
With instant setup, multi-chain support, and cloud-based IaaS, our proposed Matrix Syncer is designed to be the ultimate solution for developing complex metaverse DApps that demand advanced on-chain data operation and event reflection. All of these advantageous infrastructures can help hasten the construction of a functional metaverse that is more user-friendly, interoperable, and accessible. By using the elasticity, high availability, and flexibility of cloud computing, Matrix Syncer bridges the barrier between web 2.0 and web 3.0 by leveraging the strengths of each, and enabling the rapid development of a functional, decentralized metaverse with enhanced user-friendliness, interoperability, and accessibility. The future direction will be to integrate additional decentralized protocols into the present system because, as with the advancement of the blockchain's fundamental technology, the current web 3.0 shortcomings will undoubtedly be resolved in the future.

\bibliographystyle{IEEEbib}
\bibliography{paper}

\end{document}